\documentclass[12pt,reqno]{amsart}
\usepackage{amssymb}
\usepackage{float}
\usepackage{graphicx}
\usepackage{textcomp}
 \usepackage{xcolor}
\usepackage[colorlinks=true, linkcolor=red, citecolor=blue]{hyperref}
\theoremstyle{plain}

\numberwithin{equation}{section}

\usepackage[top=3cm,bottom=3cm,left=3cm,right=3cm]{geometry}
\begin{document}
\title{ {A \MakeLowercase{$q$-deformation of  true-polyanalytic} B\MakeLowercase{argmann transforms when $q^{-1}>1$}}}
\author{ O\MakeLowercase{thmane} E\MakeLowercase{l Moize}$^{\ast }$ \MakeLowercase{and} Z\MakeLowercase{ouhair} M\MakeLowercase{ouayn}$^{\flat }$}
\maketitle
\vspace*{-0.7em}
\begin{center}
\textit{{\footnotesize ${}^{\ast }$ Department of Mathematics, Faculty of Sciences,\\Ibn Tofa\"{i}l University, P.O. Box. 133, K\'enitra, Morocco\vspace*{0.2mm}\\[3pt]
${}^{\flat }$ Department of Mathematics, Faculty of Sciences
and Technics (M'Ghila),\vspace*{-0.2em}\\ Sultan Moulay Slimane University, P.O. Box. 523, B\'{e}ni Mellal, Morocco  }
}
\end{center}
\begin{abstract}
\scriptsize{We combine continuous  $q^{-1}$-Hermite Askey polynomials with new  $2D$ orthogonal polynomials introduced by Ismail and Zhang as $q$-analogs for complex Hermite polynomials  to construct a new set of coherent states depending on a nonnegative integer  parameter $m$. In the analytic case corresponding to $m=0$, we recover a known result on the Ar\"{\i}k-Coon oscillator for $q'=q^{-1}>1$. Our construction leads to a new $q$-deformation of the $m$-true-polyanalytic Bargmann transform on the complex plane. The obtained result may be used to introduce a $q$-deformed Ginibre-type point process.}
\end{abstract}
\section{Introduction and statement of the results}

In \cite{B}, Bargmann introduced a transform which maps isometrically
the space $L^{2}(\mathbb{R})$ onto the Fock space $\mathfrak{F}(\mathbb{C})$
of entire functions belonging to $\mathfrak{H}:=L^{2}\left( 
\mathbb{C},e^{-z\bar{z}}d\lambda(z)/\pi \right) $  where $d\lambda(z) $ is the Lebesgue measure on $\mathbb{C}$%
. Since this transform is strongly linked to the Heisenberg group, it
can be seen as a windowed Fourier transform \cite{BH}. Hence, the important
role it plays in signal processing and harmonic analysis on the phase space 
\cite{Folland}. It is also possible to interpret the kernel of this
transform in terms of coherent states \cite{AGA} of the quantum harmonic oscillator whose eigenstates are given by Hermite functions 
\begin{equation}\label{herfunc1}
\varphi_j(\xi)=\left(\sqrt{\pi} 2^j j!\right)^{-1/2}H_j(\xi)e^{-\frac{1}{2}\xi^2},
\end{equation}
$H_j(\cdot)$ being the $j$th Hermite polynomial (\cite{KS}, p.50). A coherent
state can be defined by a normalized  vector $\Psi_z $ in $L^{2}(\mathbb{R})$, as
 a special superposition with the form
\begin{equation}\label{canonCS}
\Psi_z:=\left(e^{z\bar{z}}\right)^{-1/2}\displaystyle\sum_{j\geq 0} \frac{z^j}{\sqrt{j!}}\varphi_j,\quad z \in \mathbb{C}.
\end{equation} 
It turns out that  the coefficients 
\begin{equation}\label{hjclas}
h_{j}(z):=\frac{z^{j}}{\sqrt{j!}},\ j=0,1,2,...,  
\end{equation}%
form an orthonormal basis of $\mathfrak{F}(\mathbb{C})$. If we denote by $\mathcal{B}_0$
the Bargmann transform, the image of an arbitrary function $f\in L^{2}(%
\mathbb{R})$ can be written as 
\begin{equation}\label{barcl}
\mathcal{B}_0[f](z):=\pi ^{-\tfrac{1}{4}}\int_{\mathbb{R}%
}e^{-\tfrac{1}{2}z^{2}-\tfrac{1}{2}\xi ^{2}+\sqrt{2}\xi z}f(\xi )d\xi ,\quad z\in \mathbb{C}. 
\end{equation}%
Otherwise, it was proven \cite{AIM} that $\mathfrak{F}(\mathbb{C})$ co\"{\i}ncides with the null space 
\begin{equation}
\mathcal{A}_{0}(\mathbb{C}):=\{F \in \mathfrak{H},\text{ }\tilde{\Delta}F =0\}  
\end{equation}%
of the second-order differential operator 
\begin{equation}
\tilde{\Delta}:=-\frac{\partial ^{2}}{\partial z\partial \bar{z}}+\bar{z}%
\frac{\partial }{\partial \bar{z}}. 
\end{equation}%
The latter one, which acts on the Hilbert space, can be unitarly intertwined to
appear as the Schr\"{o}dinger operator for the motion of a charged
 particle evolving in a constant and uniform magnetic field normal to the
plane. The spectrum of  $\tilde{\Delta}$ in $\mathfrak{H}$ is
the set of eigenvalues $m\in \mathbb{Z}_{+}$, each of which has an infinite multiplicity,
usually called Euclidean Landau levels. For $m\in \mathbb{Z}_{+}$, the associated eigenspace \cite{AIM} :
\begin{equation}\label{mtruespace}
\mathcal{A}_{m}(\mathbb{C}):=\left\{ F\in \mathfrak{H},\text{ }\tilde{\Delta}F=mF\right\}   
\end{equation}%
is also the $m$th-true-polyanalytic  space \cite%
{Vasilevski,AF} or the generalized Bargmann space \cite{AIM}. An orthonormal basis for this space is given by the
functions 
\begin{equation}
\label{truebase}
h_{j}^{m}(z):=(-1)^{m\wedge j}\left( m!j!\right) ^{-1/2}(m\wedge
j)!|z|^{|m-j|}e^{-i(m-j)arg(z)}L_{m\wedge j}^{(|m-j|)}(z\bar{z}),\text{ }%
j=0,1,...,  
\end{equation}%
$L_{n}^{(\alpha )}(\cdot)$ being the Laguerre polynomial (\cite{KS}, p.47), $%
m\wedge j=\mathrm{min}(m,j)$ and $i^2=-1$. Note that when  $m=0,$ $h_{j}^{0}(z)$ reduces to $h_{j}(z)$  
in \eqref{hjclas}. Therefore, we may replace the coefficients $h_{j}(z)$
by $h_{j}^{m}(z)$ to construct a family of coherent states depending on the parameter $m$.
This leads to the coherent states  transform $\mathcal{B}_{m}:L^{2}(%
\mathbb{R})\rightarrow \mathcal{A}_{m}(\mathbb{C})$, defined for any $f\in
L^{2}(\mathbb{R})$ by \cite{Mouayn1}: 
\begin{equation}\label{tr1}
\mathcal{B}_{m}[f](z)=(-1)^{m}(2^{m}m!\sqrt{\pi })^{-\tfrac{1}{2}}\int_{%
\mathbb{R}}e^{-\tfrac{1}{2}z^{2}-\tfrac{1}{2}\xi ^{2}+\sqrt{2}\xi
z}H_{m}\left( \xi -\frac{z+\bar{z}}{\sqrt{2}}\right) f(\xi )d\xi ,  
\end{equation}%
where $H_{m}(\cdot)$ denotes the Hermite polynomial. This transform, also called  $m$-true-polyanalytic
Bargmann transform, has found applications in time-frequency analysis \cite%
{Abr2010}, discrete quantum dynamics \cite{AoP} and determinantal point
processes \cite{APRT}. For more details on \eqref{tr1}, see \cite{AF} and
reference therein.\medskip 

We also observe that the coefficients \eqref{truebase} can be rewritten in terms
of $2D$ complex Hermite polynomials introduced by It\^{o} \cite{IK}, as $h_{j}^{m}(z)=\left( m!j!\right)
^{-1/2}H_{m,j}(z,\bar{z})$ where 
\begin{equation}
H_{r,s}(z,w)=\sum_{k=0}^{r\wedge s}(-1)^{k}k!\binom{r}{k}\binom{s}{k}%
z^{r-k}w^{s-k},\quad r,s=0,1,2,...\ .  
\end{equation}%
For the latter ones, Ismail and Zhang  have introduced the following 
$q$-analogs  (\cite{IZ}, p.9) :
\begin{equation}\label{Hjmq0}
H_{r,s}(z,w|q):=\sum_{k=0}^{r\wedge s}%
\begin{bmatrix}
r \\ 
k%
\end{bmatrix}%
_{q}%
\begin{bmatrix}
s \\ 
k%
\end{bmatrix}%
_{q}q^{(r-k)(s-k)}(-1)^{k}q^{\binom{k}{2}}(q;q)_{k}z^{r-k}w^{s-k},\ z,w\in 
\mathbb{C}  
\end{equation}%
where 
\begin{equation}\label{qshifdef}
\begin{bmatrix}
n \\ 
k%
\end{bmatrix}%
_{q}=\frac{(q;q)_{n}}{(q;q)_{n-k}(q;q)_{k}},\text{ }k\in\mathbb{Z}_+,\quad
(a;q)_{n}=\medskip \prod\limits_{l=0}^{n-1}\left( 1-aq^{l}\right)\;\mathrm{and}\; (a;q)_{\infty}=\medskip \prod\limits_{l=0}^{\infty}\left( 1-aq^{l}\right).
\end{equation}%
The polynomials \eqref{Hjmq0} can also be rewritten in a form similar to \eqref{truebase} as 
\begin{equation}\label{Hmj0}
H_{r,s}(z,w|q)=(-1)^{r\wedge s}(q;q)_{r\wedge
s}|z|^{|r-s|}e^{-i(r-s)arg(z)}L_{r\wedge s}^{(|r-s|)}\left(
zw;q\right)   
\end{equation}%
in terms of  $q$-Laguerre polynomials $L_{n}^{(\alpha) }(x;q)$ (\cite{KS}, p.108).\medskip 

Here, we introduce a new $q$-deformation of the transform \eqref{tr1} with the parameter range $q^{-1}>1$. The
kernel of such a transform may be obtained, up to a normalization factor
depending on $z$, as the closed form of a generalized coherent state
(a special superposition) that we now construct by replacing the coefficients $h_{j}^{m}(z)$ by
a slight modification of the polynomials $H_{m,j}(z,\bar{z}|q)$. More precisely, our superposition combines the new coefficients with continuous $q^{-1}$-Hermite Askey functions \cite{askey89}, which are chosen as $q-$analogs of eigenstates of the harmonic oscillator and may also be associated with the  Ar\"{\i}k-Coon oscillator for $q'=q^{-1}>1$ \cite{burb}. Precisely, by setting $w=\bar{z}$ in \eqref{Hmj0}, we will be concerned with the following new coefficients
\begin{equation}\label{Hmj}
\mathfrak{h}_{j}^{m,q}(z):=\frac{(-1)^{m\wedge j}(q;q)_{m\wedge j}\sqrt{q^{-1}(1-q)}^{|m-j|}|z|^{|m-j|}e^{-i(m-j)arg(z)}}{q^{\frac{-1}{4}((m-j)^{2}+m+j)%
}
\sqrt{(q;q)_{m}(q;q)_{j}}}L_{m\wedge j}^{(|m-j|)}\left( q^{-1}\alpha%
;q\right) 
\end{equation}%
where $\alpha=(1-q)z\bar{z}$. Since $\lim_{q\rightarrow 1}L_{n}^{(\alpha) }\left( (1-q)x;q\right)
=L_{n}^{(\alpha )}(x)$ it follows, after a straightforward calculations,
that $\displaystyle\lim_{q \to 1}\mathfrak{h}_{j}^{m,q}(z)=h_{j}^{m}(z)$ which justifies our
choice for the functions   \eqref{Hmj}. Next, as $q$-analogs of eigenstates of the Hamiltonian of the harmonic oscillator, we will be dealing with the functions
\begin{equation}\label{varphixi}
\varphi _{j}^{q}(\xi ):=\sqrt{\omega _{q}(\xi )}\left( \frac{q^{\frac{j(j+1)%
}{2}}}{(q;q)_{j}}\right) ^{\frac{1}{2}}{h}_{j}\left( \sqrt{\frac{1-q%
}{2}}\xi |q\right) ,\;\,\xi \in \mathbb{R},\text{ \ }j=0,1,2,\cdots , 
\end{equation}%
where ${h}_{j}(x|q)$ are the continuous $q^{-1}$-Hermite Askey polynomials  \cite{askey89} defined by
\begin{equation}\label{contiqher}
{h}_j(x|q)=i^{-j}H_j(ix|q^{-1}),
\end{equation} 
$H_j(x|p)$ being the continuous $p$-Hermite polynomial  with $p>1$ (\cite{KS}, p.115) and
\begin{equation}\label{omegaq}
\omega _{q}(\xi )=\pi ^{-\frac{1}{2}}q^{\frac{1}{8}}\cosh (\sqrt{\frac{1-q}{2%
}}\xi )e^{-\xi ^{2}} . 
\end{equation}%
Furthermore, in (\cite{ataki95}, p.5) Atakishiyev  showed that the polynomials \eqref{contiqher} satisfy a Ramanujan-type orthogonality relation on the full real line, which translates to
\begin{equation}\label{arthogphij}
\int\limits_{\mathbb{R}}\varphi _{j}^{q}(\xi )\varphi _{k}^{q}(\xi )d\xi
=\delta _{jk}
\end{equation}
in terms of the functions $\{\varphi _{j}^{q}\}$. The latter ones also  satisfy $\lim_{q\to 1}\varphi _{j}^{q}(\xi )=\varphi_j(\xi)$ where $\varphi_j(\xi)$ are the Hermite functions \eqref{herfunc1}. This justifies our choice in \eqref{varphixi}.   

Now, with the above material, we are able to define \textit{"\`{a} la Iwata"} \cite{dodo01, Iwata} a new family of \ generalized coherent states belonging to $L^2(\mathbb{R})$ by setting   
\begin{equation}\label{CS1}
\Psi _{z,m,q}:=(\mathcal{N}_{m,q}(z\bar{z}))^{-\tfrac{1}{2}}\sum_{j\geq 0}%
\overline{\mathfrak{h}_{j}^{m,q}(z)}\varphi _{j}^{q}, 
\end{equation}%
where the normalization factor 
\begin{equation}\label{normfac}
\mathcal{N}_{m,q}(z\bar{z})=\frac{q^{2m}((q-1)z\bar{z};q)_{\infty
}(q^{-1}(q-1)z\bar{z};q)_{m}}{((q-1)z\bar{z};q)_{m}}, 
\end{equation}%
is defined for every $z\in \mathbb{C}$. These states  satisfy the resolution of the identity operator on $L^2(\mathbb{R})$ as
\begin{equation}\label{resoiden}
\int_\mathbb{C} |\Psi _{z,m,q}\rangle\langle \Psi _{z,m,q}|d\nu _{m,q}(z)=\textbf{1}_{L^{2}(\mathbb{R})}.
\end{equation}
Here, the Dirac’s bra-ket notation  $|\Psi _{z,m,q}\rangle\langle \Psi _{z,m,q}|$ means the rank-one operator $\phi\longmapsto\langle \Psi _{z,m,q},\phi\rangle\cdot\Psi _{z,m,q} $, $\phi\in L^{2}(\mathbb{R})$ and  $d\nu_{m,q}(z):=\mathcal{N}_{m,q}(z\bar{z})d\mu_{q}(z)$ where  $d\mu _{q}(z)$  is one of many  orthogonal measures for the polynomials $\mathfrak{h}_j^{m,q}(z)$ and it is given by (\cite{IZ}, p.11) :
\begin{equation}\label{dmuq1}
d\mu _{q}(z):=\frac{q-1}{q\,\mathrm{Log}\, q}\,(E_{q}(q^{-1}z\bar{z}))^{-1}d\lambda(z)/\pi, 
\end{equation}%
where $E_{q}(x)=((q-1)x;q)_{\infty }$ defines a $q$-exponential function (\cite{GR}, p.11). Moreover, in the limit $q \to 1$ the measure $d\mu _{q}$ reduces to the Gaussian measure $e^{-z\bar{z}}d\lambda(z)/\pi $. Eq.\eqref{resoiden} may also be understood  in the  weak sense as
\begin{equation}\label{residenweak}
\int\limits_{\mathbb{C}}\langle f,\,\Psi _{z,m,q}\rangle\langle \Psi _{z,m,q},\,g\rangle d\nu _{m,q}(z)=\langle f,\,g \rangle,\qquad f,g\in \,L^2(\mathbb{R}).
\end{equation}
  Furthermore, straightforward calculations give the overlapping function of two coherent
states \eqref{CS1}. See Subsection \textbf{2.1} below for the proof.\medskip \newline
\textbf{Proposition 1.} For $m\in \mathbb{Z}_{+}$\textit{\ and }$q^{-1}>1$\textit{%
, the following assertion holds true} 
\begin{equation}\label{overlapping}
\langle \Psi _{z,m,q},\Psi _{w,m,q}\rangle_{L^2(\mathbb{R})} =\frac{q^{2m}((q-1)z\bar{w}%
;q)_{\infty }}{(\mathcal{N}_{m,q}(z\bar{z})\mathcal{N}_{m,q}(w\bar{w}))^{%
\tfrac{1}{2}}}\;{}_{3}\phi _{2}\left( 
\begin{array}{c}
q^{-m},q\frac{\bar{w}}{\bar{z}},q\frac{{z}}{{w}} \\ 
q,(q-1)z\bar{w}%
\end{array}%
\Big|q;q^{m-1}(q-1)w\bar{z}\right)   
\end{equation}%
\textit{for every} $z,w\in \mathbb{C}$. \medskip \newline
Here, the ${}_{3}\phi _{2}$ $q$-series is defined by (\cite{GR}, p.4) :
\begin{equation}\label{hgdefsimpl}
{}_{3}\phi _{2}\left( 
\begin{array}{c}
q^{-n},a,b \\ 
c,d%
\end{array}%
\Big|q;x\right)=\sum_{k\geq 0} \frac{(q^{-n};q)_{k}(a;q)_k(b;q)_k}{(c;q)_k(d;q)_k}\,\frac{x^k}{(q;q)_k}.
\end{equation}
In particular, for $z=w$ in \eqref{overlapping}, the condition $\langle \Psi _{z,m,q},\,\Psi _{z,m,q}\rangle_{L^{2}(\mathbb{R})}=1$ may provide us with the normalization factor \eqref{normfac}. Furthermore, \eqref{overlapping} gives an explicit expression for the  function
\begin{equation}\label{repkern}
K_{m,q}(z,w):=(\mathcal{N}_{m,q}(z\bar{z})\mathcal{N}_{m,q}(w\bar{w}))^{%
\tfrac{1}{2}}\langle \Psi _{z,m,q},\Psi _{w,m,q}\rangle_{L^2(\mathbb{R})} 
\end{equation}%
which satisfies the limit  
\begin{equation}\label{repkerlimit}
\lim_{q \to 1}K_{m,q}(z,w)=e^{z\bar{w}}L_{m}^{(0)}\left(
|z-w|^{2}\right) .  
\end{equation}%
The proof of \eqref{repkerlimit} is given in Subsection \textbf{2.2} below. Hence, one can say that the closure in $\mathfrak{H}_q:=L^{2}(\mathbb{C},d\mu
_{q})$ of the linear span of $\{\mathfrak{h}_{j}^{m,q}\}_{j\geq 0}$ \ is a Hilbert
space whose reproducing kernel is given  in \eqref{repkern} and  it will be called a generalized Ar\"{\i}k-Coon space of index 
$m$ and  denoted $\mathcal{A}_{m}^{q}(\mathbb{C})$. This space can also be
viewed as a $q$-analog of the $m$th-true-polyanalytic space $\mathcal{A}_{m}(%
\mathbb{C})$ in \eqref{mtruespace} whose reproducing kernel was given by  
  $e^{z\bar{w}}L_{m}^{(0)}\left( |z-w|^{2}\right) $, see \cite{AIM}. 
  
Eq.\eqref{residenweak} also means that the \textit{coherent states transform} $\mathcal{B}_{m}^{q}:L^{2}(\mathbb{R}%
)\longrightarrow \mathcal{A}_{m}^{q}(\mathbb{C})$ defined as usual (see \cite{AGA}, p.27 for the general theory) by 
\begin{equation}\label{CSTdef}
\mathcal{B}_{m}^{q}[f](z)=(\mathcal{N}_{m,q}(z\bar{z}))^{\tfrac{1}{2}}\langle f,\,\Psi _{z,m,q}\rangle_{L^2(\mathbb{R})},\;z\in \mathbb{C},
\end{equation}
is an isometric map for which we establish the following precise result, see Subsection \textbf{2.3} below for the proof.
\medskip\\
\textbf{Theorem 1.} \textit{For }$m\in \mathbb{Z}_{+}$\textit{\ and }$q^{-1}>1$%
\textit{, the  transform  \eqref{CSTdef} is explicitly given by}
\begin{eqnarray*}
\mathcal{B}_{m}^{q}[f](z)=\gamma_{q,m}\displaystyle\int_{\mathbb{R}}(-q^{\frac{1+m}{2}}\sqrt{1-q}ze^{\mathrm{argsinh}(\sqrt{\frac{1-q%
}{2}}\xi )},q^{\frac{1+m}{2}}\sqrt{1-q}ze^{-\mathrm{argsinh}(\sqrt{\frac{1-q%
}{2}}\xi )};q)_{\infty }
\end{eqnarray*}
\begin{equation}\label{Bmq}
\times\tilde{Q}_{m}\left( \sqrt{\frac{1-q}{2}}\xi ;iq^{%
\frac{m-1}{2}}\sqrt{1-q}z,iq^{\frac{m-3}{2}}\sqrt{1-q}\bar{z};q\right) \sqrt{%
\omega _{q}(\xi )}f(\xi )d\xi,
\end{equation}
\textit{where} $\gamma_{q,m}=\frac{(-1)^{m}q^{\frac{1}{2}\binom{m}{2}}}{\sqrt{(q;q)_{m}}}$ \textit{and} $\tilde{Q}_{m}$ \textit{denotes the} $%
q^{-1}$-\textit{AL-Salam-Chihara polynomials} .\medskip\\
Here, the  polynomial $\tilde{Q}_{m}$ is defined by (\cite{ataki95}, p.6) :
\begin{equation}\label{Alsamadef}
\tilde{Q}_n(\sinh \kappa;t,\tau;q)=q^{-\binom{n}{2}}(it)^n(it^{-1}e^{\kappa},-it^{-1}e^{-\kappa};q)_n\: {}_3 \phi_2\left(\begin{matrix}q^{-n},q^{1-n}t\tau,0\\ iq^{1-n}te^{\kappa},-iq^{1-n}te^{\kappa} \end{matrix}\Big|q;q\right) 
\end{equation}
where $\kappa\in\mathbb{R}\:\, \mathrm{and}\:\, t,\tau \in \mathbb{C}$. The isometry $\mathcal{B}_{m}^{q}$ will be called a $q$-deformation of
the true-polyanalytic Bargmann transform $\mathcal{B}_{m}$ when $q^{-1}>1$. Indeed, when 
$q \to 1$ \eqref{Bmq} reduces to  \eqref{tr1}, see Subsection \textbf{2.4} below for the proof.\medskip\\
\textbf{Corollary 1.} \textit{For $m=0$, the transform \eqref{Bmq} reduces to $\mathcal{B}_{0}^{q}:L^{2}(\mathbb{R}%
)\longrightarrow \mathcal{A}_0^{q}(\mathbb{C})$, defined by}
\begin{eqnarray*}\label{Bm0}
\mathcal{B}_{0}^{q}[f](z)=\int_{%
\mathbb{R}}\left(-\sqrt{q(1-q)}ze^{\mathrm{argsinh}(\sqrt{\frac{1-q%
}{2}}\xi )},\sqrt{q(1-q)}ze^{-\mathrm{argsinh}(\sqrt{\frac{1-q%
}{2}}\xi )};q\right)_{\infty }\sqrt{%
\omega _{q}(\xi )}f(\xi )d\xi 
\end{eqnarray*}%
\textit{for every $z\in\mathbb{C}$. In particular, when $q \to 1 $, $\mathcal{B}_{0}^q$ goes to the  Bargmann transform} \eqref{barcl}.\medskip\\
Here, $\mathcal{A}_0^{q}(\mathbb{C})$ is the completed space of  entire functions in  $\mathfrak{H}_q$, for which the elements 
\begin{equation}\label{elembasis}
\mathfrak{h}_j^{0,q}(z)=([j]_q!)^{-1/2}q^{\frac{1}{2}\binom{j}{2}}z^j,
\end{equation}
where $[j]_q!=\frac{(q;q)_j}{(1-q)^j}$, constitute an  orthonormal basis. Note that by replacing in \eqref{elembasis} the parameter $q$ by its inverse $q'=q^{-1}$, we recover   the   orthonormal basis $([j]_{q'}!)^{-1/2} z^j$ of the  Ar\"{\i}k-Coon type space with $q'=q^{-1}>1$ \cite{ACO}.\medskip\\
\textbf{Remark 1.} In (\cite{IZ}, p.4) Ismail and Zhang have also introduced another class of $2D$ orthogonal $q$-polynomials, here denoted by  $\tilde{H}_{m,j}(z,w|q)$, which also generalize the complex Hermite  polynomials \cite{IK} and are connected to the ones in \eqref{Hjmq0} by
\begin{equation}
\tilde{H}_{m,j}(z,w|q)=q^{mj} i^{m+j}H_{m,j}(z/i,w/i|q^{-1}) .
\end{equation}
In our previous joint work with Arjika \cite{SOZ18}, we have combined the  polynomials $\tilde{H}_{m,j}(z,\bar{z}|q)$ with the continuous $q$-Hermite polynomials $H_j(\xi|q)$ to obtain  a $q$-deformed  $m$-true-polyanalytic Bargmann transform on $L^2\left(]\frac{-\sqrt{2}}{\sqrt{1-q}},\frac{\sqrt{2}}{\sqrt{1-q}}[,\,d\xi\right)$ with $q^{-1}>1$.\medskip\\
\textbf{Remark 2.} For $m=0$, we recover in   $L^2\left(\mathbb{R},\sqrt{\omega_q(\xi)}\,d\xi\right)$ the state $\langle \xi|z,0,q\rangle\equiv \left(\omega_q(\xi)\right)^{-\frac{1}{2}} \Psi_{z,0,q}(\xi) $
as a coherent state for the Ar\"{\i}k–Coon oscillator with the deformation parameter  $q'=q^{-1}>1$, which was constructed  by Burban (\cite{burb}, p.5).\medskip\\
\textbf{Remark 3.} The expression \eqref{repkern} may also constitute  a starting point to construct a $q$-deformation for the determinantal point process associated with an $m$th Euclidean Landau level  or Ginibre-type point process in $\mathbb{C}$ as discussed by Shirai \cite{shirai}.
\section{Proofs}
\subsection{Proof of Proposition 1}
By \eqref{arthogphij}-\eqref{CS1}, the overlapping function of two coherent
states is given by 
\begin{eqnarray}  \label{RKdem1}
\langle \Psi _{z,m,q},\Psi _{w,m,q}\rangle_{L^2(\mathbb{R})} &=&\left(\mathcal{N}_{m,q}(z\bar{z})\mathcal{N}_{m,q}(w\bar{w})\right)^{%
-\tfrac{1}{2}}\;\sum_{j=0}^\infty \overline{\mathfrak{h}_j^{m,q}(z)} \mathfrak{h}_j^{m,q}(w)\cr
&=& \left(\mathcal{N}_{m,q}(z\bar{z})\mathcal{N}_{m,q}(w\bar{w})\right)^{%
-\tfrac{1}{2}}\;S^{(m)}.
\end{eqnarray}
Replacing  $\mathfrak{h}_j^{m,q}(z)$ by their  expressions in \eqref{Hmj}, we can write $S^{(m)}=S_{<\infty}^{(m)}+S_{\infty}^{(m)}$, where 
\begin{eqnarray*}  \label{Asplit}
S_{<\infty}^{(m)}&=& \displaystyle\sum_{j=0}^{m-1} \frac{%
(q,q;q)_{j }q^{\frac{(m-j)^2+m+j}{2}}{(q^{-1}-1)}^{m-j}(\bar{z}w)^{m-j}}{%
(q;q)_m(q;q)_j}L_{j}^{(m-j)}\left( q^{-1}\alpha;q\right)L_{j}^{(m-j)}%
\left( q^{-1}\beta;q\right)\cr &-& \displaystyle\sum_{j=0}^{m-1} 
\frac{(q,q;q)_{m}q^{\frac{(m-j)^2+m+j}{2}}{(q^{-1}-1)}^{j-m}({z}\bar{w}%
)^{j-m}}{(q;q)_m(q;q)_j}L_{m}^{(j-m)}\left( q^{-1}\alpha%
;q\right)L_{m}^{(j-m)}\left( q^{-1}\beta;q\right),
\end{eqnarray*}
and 
\begin{eqnarray*}  \label{Sinfi}
S_{\infty}^{(m)} &=&\displaystyle\sum_{j \geq 0}\frac{%
(q,q;q)_{m}q^{\frac{(m-j)^2+m+j}{2}}{(q^{-1}-1)}^{j-m}({z}\bar{w})^{j-m}%
}{(q;q)_m(q;q)_j}L_{m}^{(j-m)}\left( q^{-1}\alpha;q%
\right)L_{m}^{(j-m)}\left( q^{-1}\beta;q\right)\cr &=&\frac{q^{\frac{%
m^2+3m}{2}}(q;q)_m}{\lambda^m}\sum_{j\geq 0}\frac{q^{\binom{j}{2}%
}\left(\lambda q^{-m}\right)^j}{(q;q)_{j}}L_{m}^{(j-m)}\left(
q^{-1}\alpha;q\right)L_{m}^{(j-m)}\left( q^{-1}\beta;q\right),
\end{eqnarray*}
where $\lambda=(1-q)z\bar{w},\,\alpha=(1-q)z\bar{z}$ and $\beta=(1-q)w\bar{w}
$. Now, we apply  the relation (\cite{MoCa}, p.3) :  
\begin{equation}  \label{proplag}
L_n^{(-N)}(x;q)=(-1)^{-N} x^N\frac{(q;q)_{n-N}}{(q;q)_n} L_{n-N}^{(N)}(x;q)
\end{equation}
for  $N=j-m,\;n=j$, $x=\alpha$ in a first time and next for  $%
x=\beta$. To obtain that $\mathcal{S}_{<%
\infty}^{(m)}=0.$ For the infinite sum, we rewrite the $q$-Laguerre
polynomial as  (\cite{KS}, p.110) :
\begin{equation}  \label{qlaguerre}
L_n^{(\gamma)}(x;q)=\frac{1}{(q;q)_n}{}_2 \phi_1\left(%
\begin{matrix}
q^{-n},-x \\ 
0%
\end{matrix}%
\left|q;q^{n+\gamma+1}\right.\right)
\end{equation}
with $n=m,\,\gamma=j-m$, $x=q^{-1}\alpha$ for $%
L_{m}^{(j-m)}\left( q^{-1}\alpha;q\right)$ and $x=q^{-1}\beta$ for $%
L_{m}^{(j-m)}\left( q^{-1}\beta;q\right)$. This gives
\begin{eqnarray}  \label{1.6}
S^{(m)} &=& \frac{q^{\frac{m^2+3m}{2}}}{\lambda^m(q;q)_m}\mathsf{S}%
_q^{(m)}(\alpha;\beta)
\end{eqnarray}
where 
\begin{equation}  \label{1.7}
\mathsf{S}_q^{(m)}(\alpha;\beta) :=\sum_{j \geq 0} \frac{q^{\binom{j}{2}%
}\left(\lambda q^{-m}\right)^j}{(q;q)_{j}}\,{}_2 \phi_1\left(%
\begin{matrix}
q^{-m},-q^{-1}\alpha \\ 
0%
\end{matrix}%
\left|q;q^{j+1}\right.\right){}_2 \phi_1\left(%
\begin{matrix}
q^{-m},-q^{-1}\beta \\ 
0%
\end{matrix}%
\left|q;q^{j+1}\right.\right).
\end{equation}
Recalling (\cite{GR}, p.3) :
\begin{equation}
{}_{2}\phi _{1}\left( 
\begin{array}{c}
a,b \\ 
c
\end{array}%
\Big|q;x\right)=\sum_{k\geq 0} \frac{(a;q)_k(b;q)_k}{(c;q)_k}\,\frac{x^k}{(q;q)_k},
\end{equation}
the r.h.s of \eqref{1.7} becomes 
\begin{eqnarray}  \label{1.8}
\mathsf{S}_q^{(m)}(\alpha;\beta) &=& \sum_{j \geq 0} \frac{q^{\binom{j}{2}%
}\left(\lambda q^{-m}\right)^j}{(q;q)_{j}}\,\sum_{k \geq 0} \frac{%
(q^{-m},-q^{-1}\alpha;q)_k}{(q;q)_k}\,(q^{j+1})^k \sum_{l \geq 0} \frac{%
(q^{-m},-q^{-1}\beta;q)_l}{(q;q)_l}\,(q^{j+1})^l\cr &=& \sum_{k,l \geq 0} 
\frac{(q^{-m},-q^{-1}\alpha;q)_k\;q^k}{(q;q)_k} \: \frac{(q^{-m},-q^{-1}%
\beta;q)_l\;q^l}{(q;q)_l} \,\sum_{j \geq 0} \frac{q^{\binom{j}{2}%
}\left(q^{-m+k+l}\lambda \right)^j}{(q;q)_{j}}.
\end{eqnarray}
Now, by applying the $q$-binomial theorem (\cite{GR}, p.11): 
\begin{equation}
\sum_{n \geq 0} \frac{q^{\binom{n}{2}}}{(q;q)_n}a^n=(-a;q)_{\infty}
\end{equation}
for $a=q^{-m+k+l}\lambda$, the r.h.s of \eqref{1.8} takes the form 
\begin{equation}  \label{1.9}
\mathsf{S}_q^{(m)}(\alpha;\beta)= \sum_{k,l \geq 0} \frac{(q^{-m},-q^{-1}%
\alpha;q)_k\;q^k}{(q;q)_k} \: \frac{(q^{-m},-q^{-1}\beta;q)_l\;q^l}{(q;q)_l}
\,(-q^{-m+k+l}\lambda;q)_{\infty}.
\end{equation}
By making use of the identity (\cite{KS}, p.9) :
\begin{equation}  \label{id11}
(a;q)_{\gamma}=\frac{(a;q)_{\infty}}{(aq^{\gamma};q)_{\infty}}
\end{equation}
for the factor $(-q^{-m+k+l}\lambda;q)_{\infty}$, Eq.\eqref{1.9} transforms to  
\begin{equation}
\mathsf{S}_q^{(m)}(\alpha;\beta)= (-q^{-m}\lambda;q)_{\infty}\sum_{k \geq 0} 
\frac{(q^{-m},-q^{-1}\alpha;q)_k}{(q;q)_k}\,q^k \sum_{l \geq
0} \frac{(q^{-m},-q^{-1}\beta;q)_l}{(-q^{-m}\lambda;q)_{k+l}(q;q)_l}\,q^l.
\end{equation}
Next, by the fact that $(q^{-m}\lambda;q)_{l+k}=(q^{-m}\lambda;q)_k(q^{k-m}%
\lambda;q)_l$, it follows that 
\begin{eqnarray*}  \label{eq132}
\mathsf{S}_q^{(m)}(\alpha;\beta)&=&(-q^{-m}\lambda;q)_{\infty}\sum_{k \geq 0} \frac{(q^{-m},-q^{-1}%
\alpha;q)_k}{(-q^{-m}\lambda,q;q)_k}\,q^k\,{}_2 \phi_1\left(%
\begin{matrix}
q^{-m},-q^{-1}\beta \\ 
-q^{k-m}\lambda%
\end{matrix}%
\left|q;q\right.\right).
\end{eqnarray*}
Using the identity (\cite{GR}, p.10) : 
\begin{equation}
{}_2\phi_1\left(%
\begin{array}{c}
q^{-n}, b \\ 
c%
\end{array}%
\Big|q; q\right) =\frac{(b^{-1}c;q)_n}{(c;q)_n}b^n
\end{equation}
for $n=m$, $b=-q^{-1}\beta$ and $c=-q^{k-m}\lambda $,
leads to  
\begin{eqnarray*}
\hspace*{4em}\mathsf{S}_q^{(m)}(\alpha;\beta) &=& (-q^{-m}\lambda;q)_{\infty}\sum_{k \geq 0} 
\frac{(q^{-m},-q^{-1}\alpha;q)_k}{(-q^{-m}\lambda,q;q)_k}\,q^k\,\frac{%
(q^{k+1-m}\frac{z}{w};q)_m}{(-q^{k-m}\lambda;q)_m}(-q^{-1}\beta)^m.
\end{eqnarray*}
Applying the identity (\cite{KS}, p.9) :
\begin{equation}
(aq^n;q)_r=\frac{(a;q)_r(aq^r;q)_n}{(a;q)_n}
\end{equation}
for $r=m$, $n=k$,  $a=q^{1-m}{z}/{w}$ and  $a=-q^{-m}\lambda$ in a second time, we arrive at 
\begin{eqnarray}  \label{eq15}
\hspace*{4em}\mathsf{S}_q^{(m)}(\alpha;\beta) &=& \frac{(-q^{-1}\beta)^m(q^{1-m}\frac{{z}}{{w}}%
;q)_m(-q^{-m}\lambda;q)_{\infty}}{(-q^{-m}\lambda;q)_m}{}_3\phi_2\left(%
\begin{array}{c}
q^{-m}, -q^{-1}\alpha,q\frac{{z}}{{w}} \\ 
q^{1-m}\frac{z}{w},-\lambda
\end{array}%
\Big|q; q\right).
\end{eqnarray}
Finally, by the finite Heine transformation (\cite{GA09}, p.2):  
\begin{equation}
{}_3\phi_2\left(%
\begin{array}{c}
q^{-n},\xi,\sigma\\ 
\gamma,q^{1-n}/\tau%
\end{array}%
\Big|q; q\right)=\frac{(\xi\,\tau;q)_n}{(\tau;q)_n}\;{}_3\phi_2\left(%
\begin{array}{c}
q^{-n},\gamma/\sigma,\xi \\ 
\gamma,\xi\,\tau%
\end{array}%
\Big|q; \sigma\,\tau q^n\right)
\end{equation}
for  parameters $\xi=q\frac{z}{w},\;\sigma=-q^{-1}\alpha,\;\gamma=-%
\lambda$ and $\tau=\frac{w}{z}$, \eqref{eq15} reads 
\begin{equation}
\mathsf{S}_q^{(m)}(\alpha;\beta)=\frac{(-q^{-1}\beta)^m(q^{1-m}\frac{{z}}{{w}}%
,q;q)_m(-q^{-m}\lambda;q)_{\infty}}{(-q^{-m}\lambda,\frac{w}{z};q)_m}%
\,{}_3\phi_2\left(%
\begin{array}{c}
q^{-n},q\frac{\bar{w}}{\bar{z}},q\frac{z}{w} \\ 
q,-\lambda%
\end{array}%
\Big|q; -q^{m-1}(1-q)w\bar{z}\right).
\end{equation}
Summarizing the above calculations and taking into account the previous prefactors, we arrive at the announced result  \eqref{overlapping}.  $\square$
\subsection{Proof of the limit \eqref{repkerlimit}} 
Recalling that $E_{q}(x)=((q-1)x;q)_{\infty }$, then we get that 
\begin{equation}
\lim_{q \to 1}q^{2m}((q-1)z\bar{w}%
;q)_{\infty }=e^{z\bar{w}}.
\end{equation}
By another side, using \eqref{hgdefsimpl} together with the fact that $(q^{-n};q)_k=0, \forall k>n$, the series ${}_3\phi_2$ in  \eqref{overlapping} terminates as 
\begin{eqnarray}\label{terseri}
\sigma_{m,q}(z,w):=\sum_{k=0}^{m}\frac{(q^{-m},q\frac{\bar{w}}{\bar{z}},q\frac{{z}}{{w}};q)_k}{((q-1)z\bar{w},q;q)_k}\,\frac{\left(q^{m-1}(q-1)w\bar{z}\right)^k}{(q;q)_k}.
\end{eqnarray}
Thus, from the identity (\cite{KS}, p.10) :$$
\begin{bmatrix} \gamma\\ k \end{bmatrix}_q=(-1)^kq^{k\gamma-\binom{k}{2}}\frac{(q^{-{\gamma}};q)_k}{(q;q)_k},
$$
we, successively, have 
\begin{eqnarray}\label{limgfunc}
\lim_{q \to 1}\sigma_{m,q}(z,w) &=&\sum_{k=0}^{m}\lim_{q \to 1}\left(\frac{(q^{-m};q)_k}{(q;q)_k}\frac{(q\frac{\bar{w}}{\bar{z}},q\frac{{z}}{{w}};q)_k}{((q-1)z\bar{w};q)_k}\,\frac{(1-q)^k}{(q;q)_k}\,(-1)^k\left(q^{m-1}w\bar{z}\right)^{k}\right)\cr
&=&\sum_{k=0}^{m}\lim_{q \to 1}\left(\begin{bmatrix} m\\ k \end{bmatrix}_q q^{\binom{k}{2}-mk}\frac{(q\frac{\bar{w}}{\bar{z}},q\frac{{z}}{{w}};q)_k}{((q-1)z\bar{w};q)_k}\,\frac{\left(q^{m-1}w\bar{z}\right)^{k}}{[k]_q!}\right)\cr
&=&\sum_{k=0}^{m}\begin{pmatrix} m\\ k \end{pmatrix}(-1)^k\frac{|z-w|^{2k}}{k!}.
\end{eqnarray}
By noticing that the last  sum in \eqref{limgfunc} is the evaluation of the Laguerre polynomial $L_m^{(0)}$ at  $|z-w|^{2}$, the proof of the limit \eqref{repkerlimit} is completed. $\square$
\subsection{Proof of Theorem 1}
To apply \eqref{CSTdef}, we seek for a  closed form  for the following series
\begin{eqnarray}\label{So31}
\left(\mathcal{N}_{m,q}(z\bar{z})\right)^{\frac{1}{2}}\Psi_{z,m,q}(\xi)&=&\sum_{j\geq 0} \frac{(-1)^{m\wedge j} q^{\frac{(m-j)^2+(m+j)}{4}}(q;q)_{m \wedge j}\sqrt{q^{-1}(1-q)}^{|m-j|}}{\sqrt{(q;q)_m(q;q)_j}}\cr
 &\times&|z|^{|m-j|}e^{-i(m-j)arg(z)}\,L_{m\wedge j}^{(|m-j|)}\left(q^{-1}(1-q)z\bar{z};q\right)\,\varphi_j^q(\xi)
\end{eqnarray}
which may also be written as 
\begin{eqnarray}\label{Beq1}
\frac{(-1)^mq^{\frac{m^2+3m}{4}}\sqrt{\omega_q(\xi)(q;q)_m}}{(z\sqrt{1-q})^{m}}\eta^{m,q}(\xi,z),
\end{eqnarray}
where  
\begin{eqnarray}
\label{etat}
\eta^{m,q}(\xi,z)&=&\sum_{j\geq 0}\frac{q^{\frac{-2mj+2j^2}{4}}(\sqrt{1-q}z)^j}{(q;q)_j}\,L_m^{(j-m)}\left(q^{-1}\alpha;q\right)h_j(\sqrt{\frac{1-q}{2}}\xi|q)
\end{eqnarray}
with $\alpha=(1-q)z\bar{z}$. Next, replacing the $q$-Laguerre polynomial by its  expression \eqref{qlaguerre}, \eqref{etat} becomes 
\begin{eqnarray*}
\eta^{m,q}(\xi,z)&=& \sum_{j\geq 0} \frac{q^{\frac{-2mj+2j^2}{4}}(\sqrt{1-q}z)^j}{(q;q)_j}h_j(\sqrt{\frac{1-q}{2}}\xi|q)\,\frac{1}{(q;q)_m}{}_2 \phi_1\left(\begin{matrix}q^{-m},-q^{-1}\alpha \\ 0 \end{matrix}\left|q;q^{j+1}\right.\right)\cr
&=&\frac{1}{(q;q)_m}\sum_{j\geq 0} \frac{q^{\frac{-2mj+2j^2}{4}}(\sqrt{1-q}z)^j}{(q;q)_j}h_j(\sqrt{\frac{1-q}{2}}\xi|q)\,\sum_{k \geq 0} \frac{(q^{-m},-q^{-1}\alpha;q)_k}{(q;q)_k}\,q^{k(j+1)}
\end{eqnarray*}
\begin{equation}\label{Beq2}
=\frac{1}{(q;q)_m}\sum_{k \geq 0} \frac{(q^{-m},-q^{-1}\alpha;q)_k}{(q;q)_k}\,q^k\sum_{j\geq 0} \frac{q^{\binom{j}{2}}(q^{\frac{1-m}{2}+k}\sqrt{1-q}z)^j}{(q;q)_j}h_j(\sqrt{\frac{1-q}{2}}\xi|q).
\end{equation}
By using the generating function of the $q^{-1}$-Hermite polynomials (\cite{ataki95}, p.6) :
\begin{equation}
\sum_{n\geq 0}\frac{t^nq^{\binom{n}{2}}}{(q;q)_n}h_n(x|q)=(-te^{\theta},te^{-\theta};q)_{\infty},\quad\: \sinh\;\theta=x
\end{equation}
for the parameters $t=q^{\frac{1-m}{2}+k}\sqrt{1-q}z$ and
\begin{equation}\label{thetaarg}
\sinh\;\theta=\sqrt{\frac{1-q}{2}}\xi,
\end{equation}
the r.h.s of \eqref{Beq2} takes the form
\begin{eqnarray}
\label{sdetat}
\eta^{m,q}(\xi,z)= \frac{1}{(q;q)_m}\sum_{k \geq 0} \frac{(q^{-m},-q^{-1}\alpha;q)_k}{(q;q)_k}\,q^k(-ye^{\theta}q^k,ye^{-\theta}q^k;q)_\infty,
\end{eqnarray}
where $y=q^{\frac{1-m}{2}}\sqrt{1-q}z$. By applying \eqref{id11}, it follows that 
\begin{eqnarray}
\label{qetat}
 \eta^{m,q}(\xi,z)=\frac{(-ye^{\theta},ye^{-\theta};q)_\infty}{(q;q)_m}\sum_{k \geq 0} \frac{(q^{-m},-q^{-1}\alpha;q)_k}{(-ye^{\theta},ye^{-\theta};q)_k}\,\frac{q^k}{(q;q)_k}
\end{eqnarray}
which can also be expressed as
\begin{eqnarray}
\label{qqqetat}
 \eta^{m,q}(\xi,z)=\frac{(-ye^{\theta},ye^{-\theta};q)_\infty}{(q;q)_m}{}_3 \phi_2\left(\begin{matrix}q^{-m},-q^{-1}\alpha,0\\ -ye^{\theta},ye^{-\theta} \end{matrix}\Big|q;q\right).
\end{eqnarray}
Next, recalling the definition of the $q^{-1}$-Al-Salam-Chihara polynomials in \eqref{Alsamadef} for $\kappa=\theta,\,t=iq^{m-1}y$ and $\tau=iq^{\frac{m-3}{2}}\sqrt{1-q}\bar{z}$, \eqref{qqqetat} reads 
\begin{equation}
 \eta^{m,q}(\xi,z)=\frac{(-1)^mq^{\binom{m}{2}}(-ye^{\theta},ye^{-\theta};q)_\infty\: \tilde{Q}_m(\sinh \theta;iq^{\frac{m-1}{2}}y,iq^{\frac{m-3}{2}}\sqrt{1-q}\bar{z};q)}{(yq^{m-1})^m(q^{1-m}y^{-1}e^{\theta},-q^{1-m}y^{-1}e^{-\theta};q)_m(q;q)_m}.
\end{equation}
After  some simplifications, we arrive at the following form for the series \eqref{So31}
\begin{equation*}
\sqrt{\omega_q(\xi)}(-q^{\frac{1+m}{2}}\sqrt{1-q}ze^{\theta},q^{\frac{1+m}{2}}\sqrt{1-q}ze^{-\theta};q)_{\infty}
\end{equation*}
\begin{equation}\label{Bmqlimit}
\times\frac{(-1)^mq^{\frac{1}{2}\binom{m}{2}}}{\sqrt{(q;q)_m}} \tilde{Q}_m\left(\sqrt{\frac{1-q}{2}}\xi;iq^{\frac{m-1}{2}}\sqrt{1-q}z,iq^{\frac{m-3}{2}}\sqrt{1-q}\bar{z};q\right).
\end{equation}
This ends the proof. $\square$ 
\subsection{Proof of the limit  \eqref{Bmq}} To compute the limit of the quantity in \eqref{Bmqlimit} as $q \to 1$, we first observe that
\begin{equation}
\displaystyle\lim_{q \to 1}\sqrt{\omega_q(\xi)}=\displaystyle\lim_{q \to 1}\left(\pi ^{-\frac{1}{2}}q^{\frac{1}{8}}\cosh (\sqrt{\frac{1-q}{2%
}}\xi )e^{-\xi ^{2}}\right)^{1/2}=\pi^{-\frac{1}{4}}e^{-\xi ^{2}/{2}}.
\end{equation}
Next, we denote
\begin{equation}
{G}_q(z;\xi):= (-q^{\frac{1+m}{2}}\sqrt{1-q}ze^{\theta},q^{\frac{1+m}{2}}\sqrt{1-q}ze^{-\theta};q)_{\infty}.
\end{equation}
Then by \eqref{qshifdef}, we successively obtain
\begin{eqnarray*}\label{Dzxi}
\mathrm{Log}\,G_q(z;\xi) &=&\sum_{k\geq 0}\mathrm{Log} \left(1-q^{\frac{1+m}{2}+k}\sqrt{1-q}ze^{-\theta}+q^{\frac{1+m}{2}+k}\sqrt{1-q}ze^{\theta}-q^{m+1+2k}(1-q)z^2\right)\cr
&=& q^{\frac{1+m}{2}}\sqrt{1-q}z(e^{\theta}-e^{-\theta}) \sum_{k\geq 0} q^k-q^{m+1}(1-q)z^2 \sum_{k\geq 0} q^{2k}+ {o}(1-q)\cr
&=& q^{\frac{1+m}{2}}z(e^{\theta}-e^{-\theta})\frac{1}{\sqrt{1-q}}-q^{m+1}z^2\frac{1}{1+q}+ {o}(1-q).
\end{eqnarray*}
Thus, form \eqref{thetaarg} the last equality also reads 
\begin{equation}
\mathrm{Log}\,G_q(z;\xi)=q^{\frac{1+m}{2}}\sqrt{2} z\xi-q^{m+1}z^2\frac{1}{1+q}+{o}(1-q).
\end{equation}
Therefore,  $\displaystyle\lim_{q \to 1}G_q(z;\xi)=e^{\sqrt{2}z\xi-\frac{1}{2}z^2}$. To obtain the limit of the polynomial quantity in  \eqref{Bmqlimit} as $q \to 1$ , we recall that the $q^{-1}$-Al-Salam-Chihara polynomials can be expressed as (\cite{ataki971}, p.6) :
\begin{equation}\label{relasalherm}
\tilde{Q}_n(s;a,b|q)=q^{-\binom{n}{2}}\sum_{k=0}^n \begin{bmatrix} n\\ k \end{bmatrix}_q q^{\binom{k}{2}}(ia)^{n-k}\, h_k(s;b|q)
\end{equation}
in terms of the continuous big $q^{-1}$-Hermite polynomials. The latter ones satisfy the limit  (\cite{ataki972}, p.4) :
$$\displaystyle\lim_{q \to 1} \kappa^{-n} h_n(\kappa s;2\kappa b|q)=H_n(s+ib),$$
and from \eqref{relasalherm} we conclude that
\begin{equation}\label{finallimit}
\lim_{q \to 1} \kappa^{-n}\tilde{Q}_n(\kappa s;2i\kappa a,2i\kappa b;q)=H_n(s-a-b).
\end{equation}
By applying \eqref{finallimit} for $n=m,\;s=\xi, \,a=q^{\frac{m-1}{2}}z/\sqrt{2},b=q^{\frac{m-3}{2}}\bar{z}/\sqrt{2}$ and $\kappa=\sqrt{\frac{1-q}{2}}$, we establish the following 
$$\displaystyle\lim_{q \to 1}\frac{(-1)^mq^{\frac{1}{2}\binom{m}{2}}}{\sqrt{(q;q)_m}}\, \tilde{Q}_m\left(\sqrt{\frac{1-q}{2}}\xi;iq^{\frac{m-1}{2}}\sqrt{1-q}z,iq^{\frac{m-3}{2}}\sqrt{1-q}\bar{z};q\right)$$
$$=(-1)^m(2^{m}m!)^{-\tfrac{1}{2}}H_{m}\left( \xi -\frac{z+\bar{z}}{\sqrt{2}}\right).$$
Finally, by grouping the obtained three limits, we arrive at the assertion  \eqref{Bmq}. $\square$\medskip\\
\textbf{\large Acknowledgments.} The authors would like to thank the\textit{ Moroccan Association of Harmonic Analysis $\&$ Spectral Geometry}.


\begin{thebibliography}{99}
\bibitem{B} V. Bargmann, On a Hilbert space of analytic functions and an
associated integral transform, Part I, \textit{Commun. Pure Appl. Math.} 
\textbf{14} (1961), 174-187.

\bibitem{BH} B. C. Hall, Bounds on the Segal-Bargmann transform of $L_p$
functions,\textit{\ J. Fourier Anal. Appl.} \textbf{7} (2001), 553-569.

\bibitem{Folland} G. B. Folland, Harmonic Analyse on Phase Space, 
Princeton University Press, Annals of Mathematics Studies. \textbf{122}, 1989.

\bibitem{AGA} S. Twareq Ali, J. P. Antoine and J. P. Gazeau, Coherent
states, Wavelets and their Generalizations, second edition, Springer
Science+Business Media New York, 2014.

\bibitem{AIM} N. Askour, A. Intissar and Z. Mouayn, Espaces de Bargmann g\'{e}n%
\'{e}ralis\'{e}s et formules explicites pour leurs noyaux reproduisants, 
\textit{Compt. Rend. Acad. Sci. Paris}. \textbf{325} (1997) S\'{e}rie I,
707-712.

\bibitem{Vasilevski} N. L. Vasilevski, Poly-Fock spaces, Differential
operators and related topics, \textit{Oper. Theory, Adv. Appl}. \textbf{117} (2000), 371-386.

\bibitem{AF} L. D. Abreu\textbf{, }H. G. Feichtinger,\textbf{\ }Function
spaces of polyanalytic functions\emph{,} in Harmonic and Complex Analysis
and its Application, Birkhauser, 2014, 1-38.
 
 \bibitem{KS} R. Koekoek and R. Swarttouw, The Askey-scheme of hypergeometric
orthogonal polynomials and its q-analogues, Delft University of Technology,
Delft, 1998.

\bibitem{Mouayn1} Z. Mouayn, Coherent state transforms attached to
generalized Bargmann spaces on the complex plane. \textit{Math. Nachr}. \textbf{284} (2011), 1948-1954.
\bibitem{Abr2010} L. D. Abreu, Sampling and interpolation in Bargmann-Fock
spaces of polyanalytic functions. \textit{Appl. Comp. Harm. Anal.} \textbf{29} (2010), 287-302.
\bibitem{AoP} L. D. Abreu, P. Balazs, M. de Gosson and Z. Mouayn, Discrete
coherent states for higher Landau levels. \textit{Ann. of Phys.}\textsl{\ }%
\textbf{363} (2015), 337-353.
\bibitem{APRT} L.~D. {A}breu, J.~M. {P}ereira, J.~L. {R}omero and S.~{T}%
orquato, The Weyl-Heisenberg ensemble: hyperuniformity and higher
Landau levels. \textit{J. Stat. Mech. Theor. Exp.} (2017), 043103.

\bibitem{IK} K. It\^o, Complex multiple Wiener integral, \textit{Jap.
J. Math}. \textbf{22} (1952), 63-86.
\bibitem{IZ} M. E. H. Ismail and R. Zhang, On some $2D$ Orthogonal $q$%
-polynomials, \textit{Trans. Amer. Math. Soc.} \textbf{369} (2017),
6779-6821.



\bibitem{burb}  I. M. Burban, Arik-{C}oon oscillator with {$q>1$} in the
framework of unified {$(q;\alpha,\beta,\gamma;\nu)$}-deformation, \textit{J.
Phys. A.} \textbf{43} (2010).

\bibitem{askey89} R. A. Askey, Continuous q-Hermite polynomials when $q>1$. $%
q$-Series and Partitions, (ed. D. Stanton), \textit{IMA Math. Appl.},
Springer-Verlag, New York. 1989, 151-158.
\bibitem{ataki95} N. M. Atakishiev,  Orthogonality of the Askey-Wilson
polynomials with respect to a Ramanujan-type measure. (Russian) ; translated
from \textit{Teoret. Mat. Fiz}. \textbf{102} (1995), 23-28.
\bibitem{dodo01}
V. V. Dodonov, Nonclassical states in quantum optics: a {\textasciigrave}squeezed{\textquotesingle} review of the first 75 years. \textit{J. Opt. B: Quantum Semiclassical Opt}. \textbf{4} (2002), 1-33.
\bibitem{Iwata} 
G. Iwata, Transformation functions in the complex domain. \textit{Prog. Theor. Phys.} \textbf{6} (1951), 524-528.

\bibitem{GR} G. Gasper and M. Rahman, Basic hypergeometric series,
2nd ed., Encyclopedia of Mathematics and its Applications, vol. \textbf{96},%
\textit{\ Cambridge University Press, Cambridge,} 2004,

 \bibitem{ACO}    C. Quesne, K.A. Penson and  V.M. Tkachuk, Maths-type $q$-deformed coherent states for $q>1$. \textit{Physics Letters A}, \textbf{313} (2004),  29-36.
\bibitem{SOZ18} S. Arjika, O. El Moize and  Z. Mouayn, Une $q$-d\'eformation de
la transformation de Bargmann vraie-polyanalytique. \textit{C.
 R. Acad. Sci. Paris}. \textbf{356} (2018), 903-910.

\bibitem{shirai} T. Shirai, Ginibre-type point processes and their asymptotic behavior. \textit{J. Math. Soc. Japan}. \textbf{67} (2015), 763-787.

\bibitem{MoCa} S. G. Moreno, C. Garcia and M. Esther,
$q$-Sobolev  orthogonality of the  {$q$}-{L}aguerre polynomials $\{L_{n}^{(-N)}(\cdot,q)\}_{n=0}^{\infty}$ for positive integers $N$, 
\textit{J. Korean Math. Soc.} \textbf{48} (2011), 913-926.

 \bibitem{GA09}
     E. G. Andrews, The finite {H}eine transformation, Combinatorial number theory. (2009), 1-6.

\bibitem{ataki971}
M. K. Atakishiyeva  and  N. M. Atakishiyev,  Fourier-Gauss transforms of the Al-Salam-Chihara polynomials. \textit{J. Phys. A}. \textbf{30} (1997), 655-661.
\bibitem{ataki972}
M. K. Atakishiyeva  and  N. M. Atakishiyev, Fourier-Gauss transforms of the continuous big $q$-Hermite polynomials. \textit{J. Phys. A}. \textbf{30} (1997), 559-565.
\end{thebibliography}
\end{document}